\documentclass[oneside,letterpaper, 10pt]{article}
\usepackage[utf8]{inputenc}
\usepackage{geometry}
\usepackage{url}
\usepackage{graphicx}
\title{A Security \& Privacy Analysis \\ of US-based Contact Tracing Apps}
\author{Joydeep Mitra \\ Department of Computer Science \\ Stony Brook University \\ joydeep.mitra@stonybrook.edu}

\begin{document}

\maketitle

\begin{abstract}
    With the onset of COVID-19, governments worldwide planned to develop and deploy contact tracing (CT) apps to help speed up the contact tracing process. However, experts raised concerns about the long-term privacy and security implications of using these apps. Consequently, several proposals were made to design privacy-preserving CT apps. To this end, Google and Apple developed the Google/Apple Exposure Notification (GAEN) framework to help public health authorities develop privacy-preserving CT apps. In the United States, 26 states used the GAEN framework to develop their CT apps. In this paper, we empirically evaluate the US-based GAEN apps to determine 1) the privileges they have, 2) if the apps comply with their defined privacy policies, and 3) if they contain known vulnerabilities that can be exploited to compromise privacy. The results show that all apps violate their stated privacy policy and contain several known vulnerabilities.
\end{abstract}

\section{Introduction}

Contact tracing is one of many methods health authorities use to contain the rapid spread of COVID-19 \cite{klinkenberg2006}. However, manual contact tracing to keep track of the growing pandemic has been a challenge for health authorities due to the lack of human resources and the limitations of human memory to remember all possible contacts accurately  \cite{jiang2022}. Consequently, there have been several efforts to automate the contact tracing process by developing mobile apps that use tracking technology in mobile phones, such as GPS and Bluetooth \cite{anglemyer2020, ahmed2020}. In addition, since mobile phones are ubiquitous, they can be used to effectively and quickly identify a person's contacts when they test positive for COVID.

\subsection{Background}

\subsubsection{How do Contact Tracing Apps Work?}
Contact tracing apps work by estimating if two mobile phones, X and Y, are close to each other based on a metric defined by local health authorities (e.g., 6 feet). Y will be notified of a potential infection if X tests positive and confirms the result. An app detects proximity using Bluetooth Low Energy (BLE), Global Positioning System (GPS) technology, or a combination of the two. If two phones are nearby, then the apps exchange encounter messages, which contain, among other things, random identifiers and signal strength. Apps may also include location data in the encounter messages. Each app keeps a record of all the encounter messages it has exchanged with other apps. If one of these messages is from a phone whose user has tested positive, then the app analyzes the encounter message to calculate the level of risk based on factors the local health authorities decided. This calculation is performed either locally in the device or by a central server.        

While contact tracing apps can be useful to help expedite the process of contact tracing, experts have warned that such technologies have long-term consequences for user privacy and security since they can be easily exploited and used in malicious contexts such as mass surveillance \cite{akinbi2021, rowe2020, hassandoust2021}. Therefore, contact tracing apps' design and implementation must be carefully scrutinized before being deployed and adopted.

\subsubsection{Contact Tracing App Architectures}
The need to ensure privacy in contact tracing apps has led to several proposals of app architectures that will best protect user privacy -- centralized, decentralized, and hybrid \cite{martin2020, gao}. In a centralized architecture, a central server collects and stores personally identifiable information (PII), which is further used to generate random identifiers for the encounter messages. Further, the central server is responsible for determining the users that have been potentially exposed. Government health authorities have access to the data stored in the central server and can use it to perform advanced aggregate analysis, which can be useful for understanding trends to help inform mitigation efforts. However, this approach negatively affects user privacy since the central server is assumed to be trusted and has access to vast amounts of personal information, which unauthorized or hostile entities can potentially misuse.

On the other hand, a decentralized server has minimal data. Most of the data is maintained locally on the user's device. Each app periodically communicates with the server to download a set of random identifiers that have tested positive. The app then performs a risk analysis of its encounter messages locally to determine if it had come close to a device that tested positive. This architecture helps preserve privacy better than a centralized architecture. However, it limits access to crucial data that could be used to analyze the spread of the pandemic or to understand the effectiveness of the app. The hybrid architecture combines features of the centralized and decentralized architecture, where random identifier generation is left to the local device, whereas the server manages the risk analysis and exposure notification. 

Countries have developed and deployed apps based on a centralized architecture (e.g., Singapore's TraceTogether \cite{tracetogether2020}) and decentralized architecture (e.g., the apps based on the Google/Apple framework). Further, many apps also used location data to enable automated contact tracing (e.g., India's Arogya Setu \cite{gupta2020}). However, there is no consensus among researchers about which approach is the most feasible for effective contact tracing while minimizing privacy risks \cite{vaudenay2020, li2020}.

\subsubsection{The Google Apple Exposure Notification System}

In May 2020, Google and Apple collaborated to develop the Google Apple Exposure Notification (GAEN) framework based on the decentralized architecture to help public health authorities develop contact tracing apps \cite{googleGuide, appleGuide}. 

A GAEN app works by generating a temporary key, which changes periodically. The key is used to encrypt locally stored data and to generate a random identifier. The app embeds the identifier in encounter messages and exchanges them with other apps using BLE. When a user tests positive, a public health official uses a verification server to send the user a confirmation code. The user can use the confirmation code to upload all their recent keys (e.g., last 14 days) to a key server. Each app downloads keys periodically from the key server and compares them with a set of keys exchanged in the last few days. If there is a match, the app determines the risk of the potential exposure based on a formula pre-determined by public health authorities.

\subsection{Motivation}

In the absence of an official national contact tracing app in the United States, individual US states used the GAEN framework to develop their own contact tracing apps. To date, 26 US states have developed a GAEN app. Despite the security and privacy guarantees built into the GAEN framework, people in the US lack confidence in the apps' ability and intentions to protect their privacy \cite{altmann2020}. The Center for Disease Control and Prevention (CDC) in the US recommends that healthcare authorities should conduct a third-party assessment of contact tracing apps and make the results publicly available \cite{gao}. However, according to the technology assessment conducted by the United States Government Accountability Office (GAO) at the behest of the US Congress, most states with a contact tracing app have not conducted a third-party assessment. Moreover, the states that have conducted an evaluation, have not made the results publicly available \cite{gao}. Motivated by the CDC's recommendations and the lack of assessments of the US-based apps, in this paper we are analyzing the privacy and security of GAEN-based Android apps for contact tracing in the US. Specifically, we are asking the following research questions:

\begin{itemize}

    \item \textit{RQ1: What degree of privilege do the GAEN-based Android apps for contact tracing have?} Android apps by default have least privilege. They need to request the system or users for permission to perform privileged operations (e.g., use Bluetooth). The purpose of this question is to understand the permissions used by these apps and their privacy implications.
    
    \item \textit{RQ2: Do GAEN-based Android apps for contact tracing violate their own privacy policies?} Contact tracing apps are required to publish a privacy policy to inform users of the app's capabilities and its data sharing, storage and retention policies. The purpose of this question is to determine if the privacy policy is consistent with the behavior encoded in the app's source code.
    
    \textit{RQ3: Do GAEN-based Android apps for contact tracing contain known Android app vulnerabilities?} Android apps can have vulnerabilities that can be exploited by malicious apps available locally or remotely to compromise users' privacy. The purpose of this question is to identify if GAEN-based contact tracing apps have similar vulnerabilities.  
\end{itemize}

\section{Methodology}

In this section, we describe the apps selected, the tools chosen, and the factors we considered to answer our research questions.

\subsection{App Selection}

\begin{table*}[ht]
    \centering
    \resizebox{\textwidth}{!}{\begin{tabular}{l|l|l|l|l|l|l}
         App &  App & Package & Version & Version & App size & Download\\
         State & Name & Name & Name & Code & (in MB) & Date\\
         \hline
         Alabama & GuideSafe & gov.adph.exposurenotifications & 1.10.0 & 2764 & 6.70 & Nov 12, 2021\\
         Arizona & Covid Watch Arizona & gov.azdhs.covidwatch.android & 2.1.11 & 201011 & 3.56 & Nov 19, 2021\\
         California & CA Notify & gov.ca.covid19.exposurenotifications & minted14020 & minted14020 & 10.07 & Oct 15, 2021\\
         Colorado & CO Exposure Notifications & gov.co.cdphe.exposurenotifications & minted1000003 & 10000032 & 3.38 & Nov 19, 2021\\
         Connecticut & COVID Alert CT & gov.ct.covid19.exposurenotifications & minted141006 & 141006 & 9.94 & Nov 12, 2021\\
         Delaware & COVID Alert DE & gov.de.covidtracker & 1.0.1 & 15 & 105.55 & Dec 3, 2021\\
         DC & DC CAN & gov.dc.covid19.exposurenotifications & minted1100019 & 11000192 & 11.8 & Dec 3, 2021\\
         Guam & Guam Covid Alert & org.pathcheck.guam.bt & 1.0.10 & 1947 & 64.52 & Jan 19, 2022\\
         Hawaii & AlohaSafe Alert & org.alohasafe.alert & 1.0.15 & 41 & 64.64 & Nov 5, 2021\\
         Louisiana & COVID Defense & org.pathcheck.la.bt & 1.9.1 & 2661 & 7.37 & Feb 4, 2022 \\
         Maryland & MD COVID Alert & gov.md.covid19.exposurenotifications & minted151008 & 151008 & 10.19 & Mar 11, 2022\\
         Michigan & MI COVID Alert & gov.michigan.MiCovidExposure & 1.4 & 255 & 3.19 & Oct 15, 2021\\
         Minnesota & COVIDaware MN & org.pathcheck.covidsafepathsBt.mn & 1.17.12 & 3503 & 3.19 & Mar 11, 2022\\
         Nevada & COVID Trace Nevada & gov.nv.dhhs.en & minted1200005 & 12000052 & 12.12 & Mar 25, 2022\\
         New Jersey & COVID Alert NJ & com.nj.gov.covidalert & 1.0.1 & 20 & 105.62 & Nov 5, 2021\\
         New Mexico & NM Notify & gov.nm.covid19.exposurenotifications & minted1200004 & 12000042 & 3.9 & Mar 4, 2022\\
         New York & COVID Alert NY & gov.ny.health.proximity & 1.1.5 & 81 & 105.90 & Oct 8, 2021\\
         North Carolina & SlowCOVIDNC & gov.nc.dhhs.exposurenotification & 1.6 & 205 & 3.1 & Nov 5, 2021\\
         North Dakota & Care19 Alert & com.proudcrowd.exposure & 1.2 & 10 & 7.23 & Nov 12, 2021\\
         \& Wyoming & & & & & & \\
         Pennsylvania & COVID Alert PA & gov.pa.covidtracker & 2.0.0 & 46 & 105.68 & Oct 14, 2021\\
         Utah & UT Exposure Notifications & gov.ut.covid19.exposurenotifications & minted1100011 & 11000112 & 11.82 & Dec 3, 2021\\
         Virginia & COVIDWISE & giv.vdh.exposurenotification & 1.5 & 160 & 9.32 & Mar 24, 2022\\
         Washington & WA Notify & gov.wa.doh.exposurenotifications & minted142004 & 142004 & 10.36 & Nov 5, 2021\\
         Wisconsin & WI Exposure Notification & gov.wi.covid19.exposurenotifications & minted141003 & 141003 & 9.86 & Mar 19, 2022\\
    \end{tabular}}
    \vspace{2mm}
    \caption{US-based contact tracing GAEN apps.}
    \label{tab:all_apps_info}
\end{table*}

We selected the official contact tracing apps in Android of all US states developed using the exposure notification APIs by Google and Apple (GAEN). We considered these apps since the focus of our study is to examine GAEN apps based in the US. 

We found the apps from the Android Developer's official page \cite{googleApps}. Each US-based app has a link to Google Play. We used the links to download the corresponding APK file from Google Play in an Emulator running an Android version supported by the app. We transferred the apk files from the emulator to a computer where we could reverse engineer and statically analyze the apps. 

All US states did not develop a GAEN app. Further, North Dakota and Wyoming use a single app. The Massachusetts app is built into the device and can only be installed from the device's settings, not from Google Play. We could not obtain this app since we were using an emulator to install the apps. Consequently, we did not consider the Massachusetts app in our assessment. In total, we ended up with 24 apps. Table \ref{tab:all_apps_info} lists all the selected apps along with their name, package, version information, size of the APK files, and when we downloaded them.

\subsection{Tool Selection}

Research in mobile app security and privacy has led to the development of several tools and techniques to detect vulnerabilities and malicious behavior \cite{Sufatrio2015, li2017}. The tools are based on static and dynamic analysis. Most use static analysis to either flag vulnerabilities or malicious behavior or guide subsequent dynamic analysis. Few tools use only dynamic analysis. Prior research efforts studying the efficacy of such tools have observed that for freely available tools, static analysis tools detect more known Android app vulnerabilities than dynamic analysis tools \cite{ranganath2020}. Furthermore, among the static analysis tools, MobSF \cite{mobsf} detects the most known vulnerabilities. Based on this observation, we used MobSF as the primary tool for analysis. 

\begin{figure*}
    \centering
    \resizebox{\textwidth}{!}{\includegraphics[]{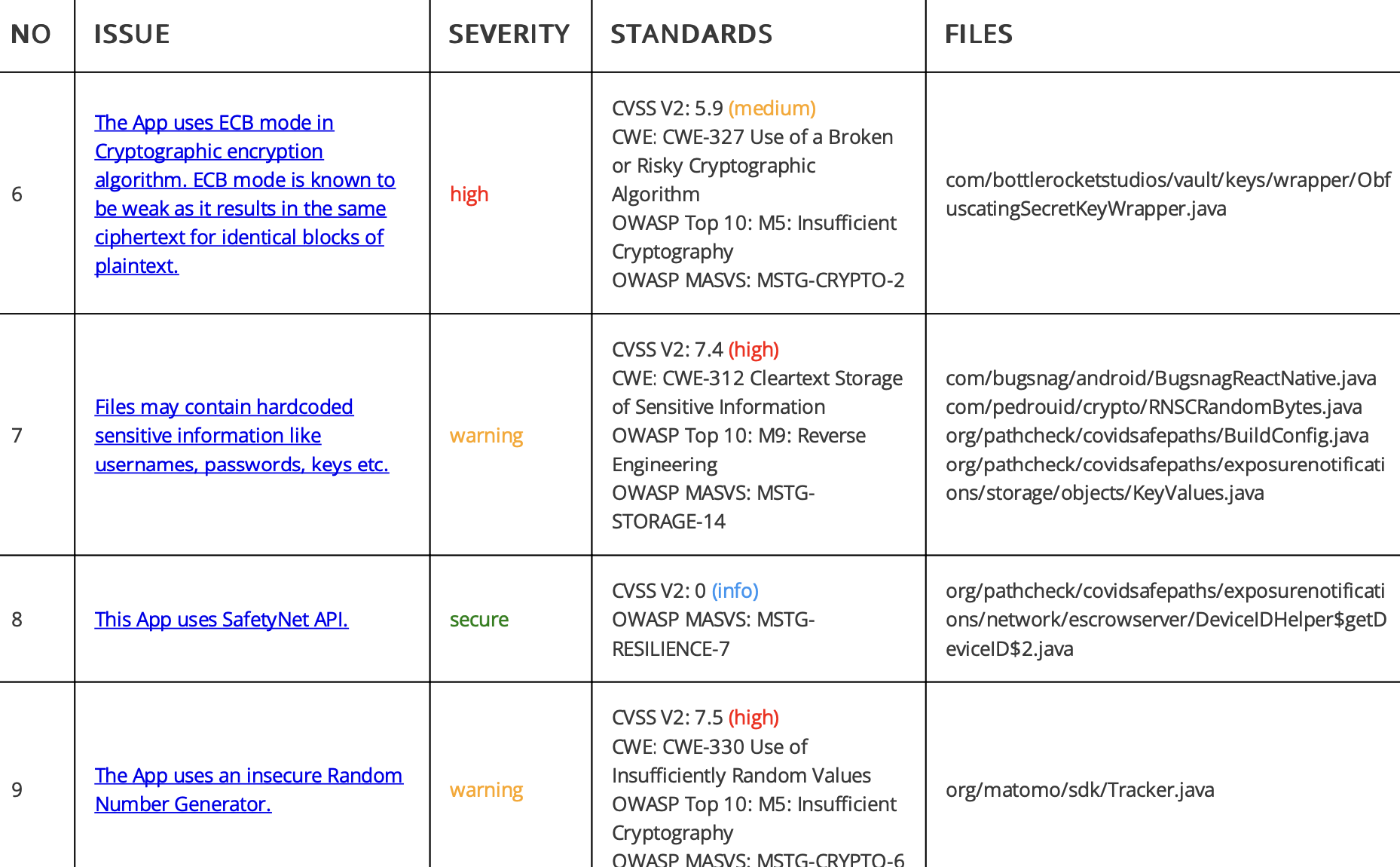}}
    \caption{Snippet from a report generated by MobSF. Each issue has a severity, security standards it is associated with, and the source file/s in which it was detected }
    \label{fig:mobsf_rpt}
\end{figure*}

MobSF has a static and dynamic analyzer. The static analyzer statically analyzes the app's source code to determine if the app uses APIs and features that are known to cause Android app vulnerabilities (see Figure \ref{fig:mobsf_rpt}). We failed to run the dynamic analyzer on the contact tracing apps that we had selected. Hence, we considered only the static analyzer.

\begin{figure*}[h]
    \centering
    \resizebox{\textwidth}{!}{\includegraphics[]{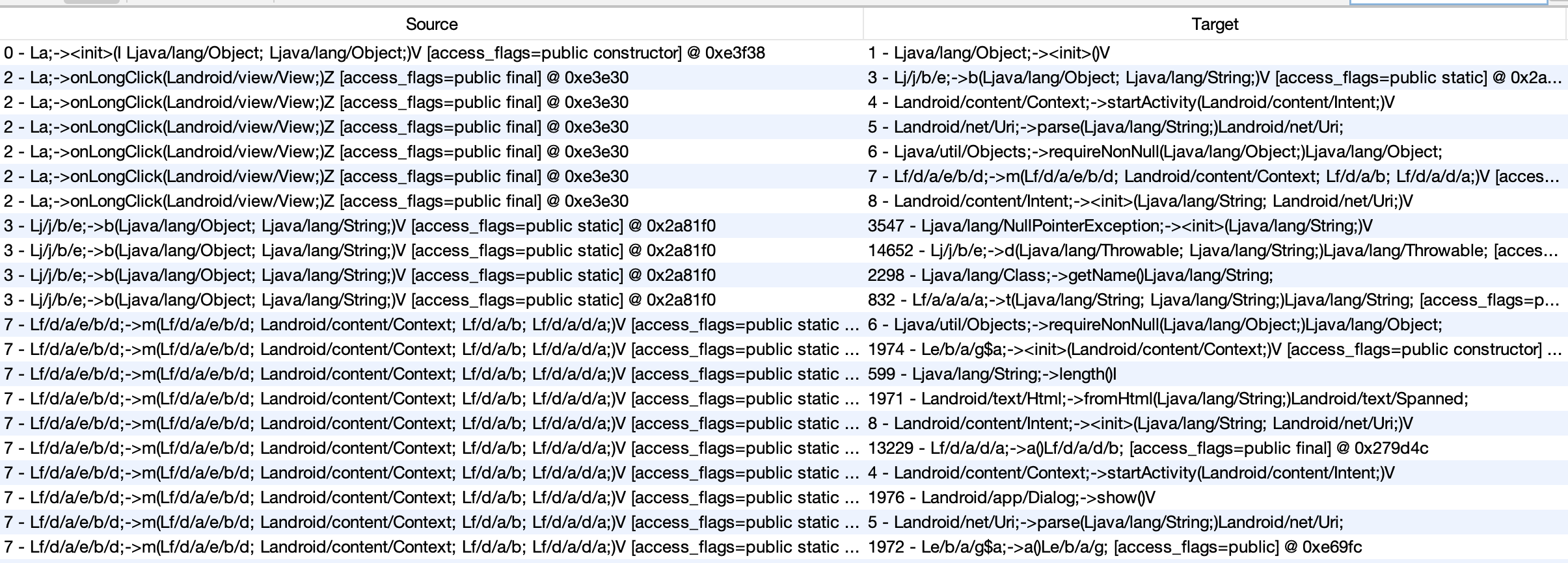}}
    \caption{Snippet of the list of edges in a control flow graph generated by the AndroGuard tool for the California app. The source and target columns indicate the nodes in the graph. Each row is a directed edge connecting the source node to the target node. Each node has an ID and an API call contained in the source code.}
    \label{fig:cfg_edges}
\end{figure*}

We used Androguard \cite{androguard} to generate the control flow graph of an app (see Figure \ref{fig:cfg_edges}), which was used to analyze the data flow through the app. We tracked the data flow in the control flow graph to determine if data flowing out of the app is sensitive or if the data being used by the app is potentially malicious. This was necessary to verify the potential data leak and data injection vulnerabilities reported by MobSF’s static analyzer, as it is known to report false positives. Specifically, we used the following strategies:
\begin{itemize}
    \item We confirmed a potential data leak vulnerability if there was a path in the control flow graph from a pre-identified sensitive data source node to a target node with shared storage, network, or inter-app communication APIs. Sensitive data sources include APIs used to collect user input (e.g., biometric), read from app’s private files and communication channels (e.g., Bluetooth), and strings hard coded with personal information (e.g., IP address). 
    
    \item We confirmed a data injection vulnerability if there was a path in the control flow graph from a pre-identified source node of potentially malicious data such as shared storage, network, or inter-app communication APIs to a target node and the target node was using the data without sanitizing it. For example, the target node is a function that uses potentially malicious data from a broadcast receiver that is exported without restrictions. 
\end{itemize}

\subsection{Policy Analysis}

A focus of our study is to determine if an app's source code is consistent with the app's privacy policy. To this end, we downloaded each app's privacy policy and examined them. We discovered that the privacy policy of an app makes the following claims:

\begin{enumerate}
    \item does not collect, store, or transmit any personally identifiable data.
    \item stores exposure data (e.g., random IDs and exposure date) locally in the users’ device.
    \item prevents unauthorized access to locally stored data.
    \item encrypts locally stored data.
    \item communicates with trusted servers through encrypted networks in the United States. 
\end{enumerate}

We used MobSF and Androguard to identify APIs or features that violate the claims listed above. If at least one such feature was found in the app, then we deemed that the app violates its defined privacy policy. For example, using MobSF, we determined instances in an app's source code where data was being stored in external storage. We then used the app's control flow graph (generated by Androguard) to determine the source of the data stored in external storage and whether the source is sensitive. If sensitive data was being stored in external storage, then we deemed it as a violation of the app's privacy policy due to bullet three listed above.

\subsection{Known Vulnerabilities Analysis}

Android apps have vulnerabilities, which malicious apps exploit to cause harm to the user \cite{lu2012, watanabe2017, mitra2017, ghafari2017}. Therefore, it is necessary to ensure such vulnerabilities do not occur in apps, especially contact tracing apps, which deal with sensitive personal information and can perform privileged operations on the phone. Hence, we analyzed these apps for known Android app vulnerabilities. 

We used MobSF, and the Ghera repository \cite{mitra2017} for our analysis. MobSF provides a list of potential vulnerabilities in its static analysis report. We investigated each of them to determine their veracity as MobSF is known to report false positives. However, MobSF does not detect all known vulnerabilities. Therefore, we used Ghera, a repository of 60 known vulnerability benchmarks, to further guide our analysis. Each benchmark in Ghera is well-documented and contains only the features related to the vulnerability captured in the benchmark. Next, we determined the features/APIs used in the apps using the static analyzer in MobSF. We then considered the app features also used in the Ghera benchmarks. We investigated each such feature to determine if it resulted in a potentially exploitable vulnerability, that is, it could be exploited by a malicious app on the device or remotely. For example, an app can give unrestricted access to another app using the pending intent feature in Android. If we found an app using pending intent, we investigated it to confirm the same, which, if true, could result in a potential privilege escalation attack.

\section{Results}
In this section, we report the findings of our study in terms of the permissions requested and used by the apps, their potential privacy violations, and the potential vulnerabilities in them that can be exploited to cause harm to the user.

\subsection{RQ1: What degree of privilege do the GAEN-based Android apps for contact tracing have?}

\begin{table*}[ht]
    \centering
    \resizebox{\textwidth}{!}{\begin{tabular}{l|l|l|l|l|l|l|l|l|l|l|l|l|r}
          & \multicolumn{12}{c}{Permissions} &\\
         \hline
         App & P1 & P2 & P3 & P4 & P5 & P6 & P7 & P8 & P9 & P10 & P11 & P12 & \# Permissions per app\\
         \hline
         Alabama & Y & Y & Y & Y & Y & Y & Y & Y & N & N & N & N & 8\\
         Arizona & Y & N & Y & Y & Y & N & Y & Y & Y & Y & N & N & 8\\
         California & Y & N & Y & Y & Y & N & Y & Y & N & N & N & N & 6\\
         Colorado & Y & N & Y & Y & Y & N & Y & Y & N & N & N & N & 6\\
         Connecticut & Y & N & Y & Y & Y & N & Y & Y & N & N & N & N & 6 \\
         Delaware & Y & Y & Y & Y & Y & Y & N & Y & N & N & N & N & 7\\
         DC & Y & N & Y & Y & Y & N & Y & Y & N & N & N & N & 6\\
         Guam & Y & Y & Y & Y & Y & Y & Y & Y & N & N & N & N & 8 \\
         Hawaii & Y & Y & Y & Y & Y & Y & Y & Y & N & N & N & N & 8\\
         Louisiana & Y & Y & Y & Y & Y & Y & Y & Y & N & N & N & N & 8 \\
         Maryland & Y & N & Y & Y & Y & N & Y & Y & N & N & N & N & 6\\
         Michigan & Y & N & Y & Y & Y & N & Y & Y & N & N & N & N & 6\\
         Minnesota & Y & Y & Y & Y & Y & Y & Y & Y & N & N & N & N & 8\\
         Nevada & Y & N & Y & Y & Y & N & Y & Y & N & N & Y & Y & 8\\
         New Jersey & Y & Y & Y & Y & Y & Y & Y & Y & N & N & N & N & 8 \\
         New Mexico & Y & N & Y & Y & Y & N & Y & Y & N & N & Y & Y & 8\\
         New York & Y & Y & Y & Y & Y & N & Y & Y & N & N & N & N & 7\\
         North Carolina & Y & N & Y & Y & Y & N & Y & Y & N & N & N & N & 6\\
         North Dakota & Y & N & Y & Y & Y & N & Y & Y & N & N & Y & Y & 8\\
         \& Wyoming &  &  &  &  &  &  &  &  &  &  &  & \\
         Penn & Y & Y & Y & Y & Y & N & Y & Y & N & N & N & N & 7\\
         Utah & Y & N & Y & Y & Y & N & Y & Y & N & N & N & N & 6\\
         Virginia & Y & N & Y & Y & Y & N & Y & Y & N & N & N & N & 6 \\
         Washington & Y & N & Y & Y & Y & N & Y & Y & N & N & N & N & 6\\
         Wisconsin & Y & N & Y & Y & Y & N & Y & Y & N & N & N & N & 6\\
         \hline
         \# Apps per & 24 & 9 & 24 & 24 & 24 & 7 & 23 & 24 & 2 & 2 & 2 & 2 &\\
         permission & \multicolumn{13}{c}{}\\
    \end{tabular}}
    \vspace{2mm}
    \caption{Permissions used by the US-based GAEN apps as declared in their manifest file. The permissions names are encoded as $P_N$ due to lack of space. The exact permission names are listed in Table \ref{tab:perm_names}}
    \label{tab:perm_info}
\end{table*}

\begin{table*}[]
    \centering
    \begin{tabular}{l|l}
         Permission Code &  Permission Name\\
         \hline
         P1 & android.permission.INTERNET \\ 
         P2 & android.permission.VIBRATE \\
         P3 & android.permission.RECEIVE\_BOOT\_COMPLETED \\
         P4 & android.permission.BLUETOOTH \\
         P5 & android.permission.ACCESS\_NETWORK\_STATE \\
         P6 & android.permission.ACCESS\_WIFI\_STATE \\
         P7 & android.permission.WAKE\_LOCK \\
         P8 & android.permission.FOREGROUND\_SERVICE \\
         P9 & com.google.android.c2dm.permission.RECEIVE \\
         P10 & com.google.android.finsky.permission.BIND\_GET\_INSTALL\_REFERRER\_SERVICE \\
         P11 & android.permission.USE\_BIOMETRIC \\
         P12 & android.permission.USE\_FINGERPRINT\\
    \end{tabular}
    \vspace{2mm}
    \caption{Name of permissions used in all US-based GAEN apps.}
    \label{tab:perm_names}
\end{table*}

A total of eight permissions are used across all 24 contact tracing apps. An app uses approximately seven permissions on average. This is higher than the average permissions used by apps in general, which is five \cite{pew2015permissions}. Furthermore, five of the eight permissions are used by all 24 apps, which suggests that only five of them are necessary for contact tracing. The other three permissions are most likely extraneous. Therefore, \textit{the GAEN apps in Android are over-privileged, which is concerning since these apps have access to vast amounts of personal data}. Over-privileged apps increase the risk of being exploited since they expose a larger attack surface due to having more privileges than needed \cite{sarma2012}. Therefore, developers of these apps should carefully consider the required permissions and ensure that only the necessary ones are used by their apps.

All permissions, except two, are normal permissions \cite{androidPerms}, that is, granted by Android when a user installs the app. Therefore, the apps do not need to ask the user for permission at runtime; they always have them. However, in the context of contact tracing, where many users have to install these apps to help prevent the spread of the pandemic, this permission model compromises privacy as it requires users to trust the apps during installation. Therefore, users have less control over granting or denying app permission. Consequently, this further hinders an app's adoption, especially in an environment where government-deployed contact tracing apps are received with skepticism \cite{prakash2022}. However, app developers cannot resolve this issue as Android, not the apps, define these permissions. Therefore, \textit{platform developers should either consider changing the way these permissions are granted or define custom permissions that should be used in the context of contact tracing}. 

Google-based third-party analytics libraries, not the core Android system, define two permissions. Only two of the 24 considered apps use these permissions. Therefore, this raises the question if these permissions are necessary for contact tracing. Further, such apps are designed to have the necessary permissions to access privileged operations in the device (e.g., Bluetooth). In this context, \textit{should contact tracing apps further increase security and privacy risks by using third-party libraries that are not directly related to the task of contact tracing?}

The \textit{USE\_FINGERPRINT} permission is deprecated \cite{fingerprintPerms}. Two of 24 apps use it. While using deprecated permissions is not recommended, this is not a major concern as both the apps use the permission in conjunction with the \textit{USE\_BIOMETRIC} permission, which is the recommended permission to use instead of \textit{USE\_FINGERPRINT}. Nevertheless, \textit{apps should not use deprecated permissions since they may have unknown and unexpected security and privacy implications}.

Two of the 24 apps use biometric permissions to access the device capabilities to use and collect biometric information. While this is not a violation of privacy by itself, it increases the risk of exposing private sensitive user info to unauthorized entities. Furthermore, \textit{since a majority of the GAEN apps are not using biometric permissions, it raises questions about the necessity of using such capabilities to collect and transmit biometric-related data}.

The \textit{ACCESS\_WIFI\_STATE} permission is used by 17 of the 24 selected apps. This suggests that not all apps need this permission for contact tracing. Apps use this permission when they connect to a remote server through WiFi. \textit{Using WiFi is not always secure since WiFi networks may not be protected and may be susceptible to Man-In-The-Middle attacks. Therefore, apps that collect and transmit sensitive information over the internet should avoid WiFi communication}.

The \textit{VIBRATE} permission is used by nine of 24 apps to control the device's vibration. This permission is not necessarily benign. If used incorrectly or maliciously, it may damage a user's phone \cite{pew2015permissions}. Therefore, it is best to avoid using such permissions if not absolutely necessary. \textit{Since a majority of the selected apps do not use the VIBRATE permission, this permission is not likely necessary for GAEN apps.}

Only one of the 24 apps declares a query element in its manifest file to indicate the list of apps it can communicate with. Specifically, the Arizona app declares that the query elements \textit{android.intent.action.DIAL} and \textit{android.intent.action.SEND} in its manifest files. This implies that the Arizona app has capabilities to place phone calls and share data with any app with the SEND intent. \textit{In the context of contact tracing, these capabilities are unnecessary and pose additional risk to the security and privacy of the app's users}.

\subsection{RQ2: Do GAEN-based Android apps for contact tracing violate their own privacy policies?}

\begin{table*}[ht]
    \centering
    \resizebox{\textwidth}{!}{\begin{tabular}{l|l|l|l|l|l|l}
          & \multicolumn{5}{c}{Privacy Policy Violation} & \\
         \hline
         App & Collects Location & Uses Insecure & Uses Weak & Uses HTTP & Communicates With Non-US & \# Violations\\
         &  & Storage & Encryption & & server domain & per app\\
         \hline
         Alabama & Y & Y & Y & N & Y & 4\\
         Arizona & Y & N & N & N & N & 1\\
         California & Y & N & N & Y & Y & 3\\
         Colorado & Y & N & N & Y & Y & 3\\
         Connecticut & Y & N & N & N & Y & 2\\
         Delaware & Y & Y & N & Y & N & 3\\
         DC & Y & N & N & Y & Y & 3\\
         Guam & N & N & Y & N & N & 1\\
         Hawaii & Y & Y & Y & N & N & 3\\
         Louisiana & Y & N & Y & N & N & 2\\
         Maryland & Y & N & N & Y & Y & 3\\
         Michigan & Y & N & N & Y & N & 2\\
         Minnesota & N & Y & Y & Y & Y & 4\\
         Nevada & Y & N & N & Y & Y & 3\\
         New Jersey & N & Y & N & Y & N & 2\\
         New Mexico & Y & Y & N & Y & Y & 4\\
         New York & N & Y & N & N & N & 1\\
         North Carolina & Y & N & Y & Y & Y & 4\\
         North Dakota & Y & Y & N & N & Y & 3\\
         Penn & Y & Y & N & N & N & 2\\
         Utah & Y & N & N & Y & N & 2\\
         Virginia & Y & N & N & N & N & 1\\
         Washington & Y & N & N & Y & Y & 3\\
         Wisconsin & Y & N & N & Y & Y & 3\\
         \hline
         \# Apps per & 20 & 9 & 6 & 14 & 13\\
         violation & & & & & \\
    \end{tabular}}
    \vspace{2mm}
    \caption{Privacy Violations by each US-based GAEN app. Columns 2-6 indicate a feature or an action that an app claims it does not use or do in its privacy policy. The cells with Y/N denote Yes if an app performs the action in the corresponding column and No if it does not. Y implies a privacy violation and N implies otherwise. }
    \label{tab:privacy_violations}
\end{table*}

All US-based GAEN apps violate their privacy policy since their behavior is inconsistent with at least one of the claims in their policy. This is concerning since contact tracing apps collect and store vast amounts of private data, which can be potentially misused. Therefore, they should take additional care to guarantee their users' privacy or at least be consistent with their own policies. 

The GAEN apps are designed not to collect, store, or track location. However, 20 of the 24 GAEN apps collect users’ locations despite claiming otherwise in their privacy policy. On further analysis, we discovered that these apps are not explicitly collecting location. However, they use a library called \textit{TwilightManager} that collects user location to determine the local time. Apps that have configured dark themes automatically import and use this library. Therefore, these apps violate their privacy policy due to using a library that collects location. Consequently, \textit{this raises the question if app developers are aware of the privacy implications of the libraries they used in their apps}. Vetting the libraries before using them is especially crucial for contact tracing apps, which malicious actors can potentially misuse to compromise user privacy.     

The Nevada and New Mexico apps collect Biometrics for authentication or to encrypt locally stored data. However, the privacy policy of these apps does not explicitly state that they collect biometric information. Moreover, they mention that the apps do not collect any personally identifiable information. Hence, we deem these apps as violating their own privacy policy.    

All apps claim in their privacy policy that they store exposure-related data in local storage in a way that prevents unauthorized access. However, nine of the 24 apps store their data in external storage. Any app installed on the device (including malicious apps) can access this data if they have the necessary permission to access external storage. Therefore, all other apps can potentially access the exposure-related data stored by these nine apps. Consequently, this leads to a violation of privacy as defined in the apps’ privacy policy. This issue could have been addressed by using the app’s internal storage instead of the external storage because, in Android, the internal storage of an app can only be accessed by the app. Moreover, Android recommends that an app’s data should be stored in internal storage unless it needs to be shared with other apps. \textit{Considering more than a third of the selected apps used external storage instead of internal storage suggests that several developers of these apps are not aware of the difference between the two. This lack of knowledge is concerning since the apps collect and store vast amounts of personal information. If access to the data is not minimized, then they can be potential targets for cyberattacks by malicious actors.} 

The privacy policy of all the selected apps mentions that data stored locally in the device is protected by encryption. However, six of the 24 apps use the AES block cipher in ECB mode, which is a weak cipher. Weak encryption is a violation of the apps’ privacy policy of encrypting local data since it leads to a potential leak of sensitive data. \textit{This result implies that a significant number of developers are not aware that AES in ECB mode should be avoided despite several security guidelines, such as OWASP, recommending not to use it.} One likely reason that developers end up using the ECB mode is that this is the default mode for AES encryption in Android. As a result, developers must explicitly change the mode. However, this oversight is surprising since the focus of the GAEN apps was to ensure user privacy by storing personally identifiable information locally and protecting them via strong encryption. \textit{Failure to use strong encryption despite claiming to do so in their privacy policies raises questions about their diligence in protecting users’ data from misuse. }

Fourteen of the 24 apps use HTTP to communicate with remote servers. Consequently, their communication can be potentially hijacked by Man-In-The-Middle attacks. Furthermore, this is inconsistent with the apps’ stated privacy policy that mentions that communication with remote servers is encrypted end-to-end. Considering that HTTP is used in more than half of the selected apps implies that the developers of these apps are either unaware of the implications of using HTTP or did not do due diligence to verify that all communication with remote web servers uses HTTPS. \textit{The use of HTTP is concerning, especially since using HTTPS is an essential requirement of apps that transmit sensitive personal information to remote web servers. Moreover, the developers of apps promoted for large-scale use and with access to vast amounts of personal information must be more perceptive of such issues and take extra care to avoid them.}  

Exposure notification apps transmit exposure-related data (e.g., random IDs and exposure date) to remote servers when a user consents to share their data in the event of an exposure to COVID-19. The apps' privacy policy does not mention the location of the servers. Therefore, users reasonably assume that the servers are in the United States. However, 13 of the 24 apps communicate with exposure notification servers outside the United States. Furthermore, the privacy policy of two of the 13 apps explicitly states that the apps only communicate with servers in the United States. \textit{Therefore, this raises the question if the apps are sending exposure-related data of US-based residents outside the US. The apps should be more transparent and explicitly state in their policies the location of the exposure notification servers so users can make informed decisions.} 

\subsection{RQ3: Do GAEN-based Android apps for contact tracing contain known Android app vulnerabilities?}

\begin{table*}[ht]
    \centering
    \resizebox{\textwidth}{!}{\begin{tabular}{l|l|l|l|l|l|l|l}
          & \multicolumn{6}{c}{Known Vulnerabilities} & \\
         \hline
         App & Unprotected & Insecure & Weak & Allows & Insecure & No Certificate & \# Vulns. \\
         & Component & PRNG & hashing & Data Backup & SSL Impl. & Pinning & per app\\
         \hline
         Alabama & Y & Y & N & N & N & Y & 3\\
         Arizona & N & Y & Y & N & Y & N & 3\\
         California & N & Y & N & N & N & Y & 2\\
         Colorado & N & Y & N & N & N & Y & 2\\
         Connecticut & Y & Y & N & N & N & Y & 3\\
         Delaware & N & Y & Y & N & N & N & 2\\
         DC & N & Y & N & N & Y & Y & 3\\
         Guam & Y & Y & Y & N & N & Y & 4\\
         Hawaii & Y & Y & N & N & N & Y & 3\\
         Louisiana & Y & Y & N & N & N & Y & 3\\
         Maryland & Y & Y & N & N & N & Y & 3\\
         Michigan & N & N & N & N & N & Y & 1\\
         Minnesota & Y & Y & N & N & N & Y & 3\\
         Nevada & Y & Y & N & Y & N & Y & 4\\
         New Jersey & N & Y & Y & N & N & N & 2\\
         New Mexico & Y & Y & N & Y & N & Y & 4\\
         New York & N & Y & Y & N & N & N & 2\\
         North Carolina & N & Y & N & N & N & Y & 2\\
         North Dakota/Wyoming & N & Y & Y & N & N & N & 2\\
         Penn & N & Y & Y & N & N & N & 2\\
         Utah & N & Y & N & Y & N & Y & 3\\
         Virginia & Y & N & N & N & N & Y & 2\\
         Washington & Y & Y & N & N & N & Y & 3\\
         Wisconsin & Y & Y & N & N & N & Y & 3\\
         \hline
         \# Apps per & 12 & 22 & 7 & 4 & 1 & 18 & \\
         Vuln & & & & & & &\\
    \end{tabular}}
    \vspace{2mm}
    \caption{Known Vulnerabilities in each US-based GAEN app. Columns 2-5 indicate a known vulnerability. The cells with Y/N denote Yes if an app contains the vulnerability and No otherwise.}
    \label{tab:known_vulns}
\end{table*}

We discovered a total of six known vulnerabilities across all 24 apps, and each app had approximately two vulnerabilities on average. All 24 apps had at least one of the six vulnerabilities. Table \ref{tab:known_vulns} shows the breakdown of the vulnerabilities found in each app. The vulnerabilities are briefly described as follows:

\begin{itemize}
    \item \textit{Unprotected Component.} Android apps consist of components. Apps can export their components to share operations or data with other apps. However, if components are exported without restrictions, all apps, including malicious apps, can access them. Consequently, it can lead to denial-of-service, data leak, and data injection attacks.
    
    \item \textit{Insecure PRNG.} Apps that use the \textit{Random} package in Java to generate pseudo-random numbers can be more easily predicted than apps that use the \textit{SecureRandom} package. Since contact tracing apps based on GAEN rely on random identifiers, they should use a random number generator that makes it harder to predict the random numbers. Therefore, apps should use the \textit{SecureRandom} package instead of the \textit{Random} package to generate hard-to-predict random identifiers.
    
    \item \textit{Weak hashing.} Hashing algorithms such as MD5 and SHA-1 are considered weak. Attackers can use a hash collision to forge a duplicate hash. Therefore, apps should avoid using them to prevent forgery attacks.
    
    \item \textit{Data Backup.} Android allows users to create backups of all data in an app without having root privileges. Consequently, malicious users with access to the device will be able to create a backup of all the app's data using a USB. Apps can be configured to protect against this potential attack by setting the \textit{allowBackup} attribute in the app's manifest file to false. Android recommends apps to disable this feature to prevent malicious users from accessing an app’s local data.
    
    \item \textit{Insecure TLS/SSL Implementation.} Apps using TLS/SSL protocol to communicate with remote servers must verify the trustworthiness of the servers. The established way to verify trust is for the app to maintain a list of trusted certificate authorities (CAs). The server is configured with a certificate containing a public key and a matching private key. The certificate must be signed by a certificate authority (CA) trusted by the app. Generally, the list of trusted CAs is pre-installed in the device on which the app is installed. However, the connection may fail if the certificate used to configure the server (1) is signed by a CA, not in the list of trusted CAs, (2) or is self-signed, (3) or is signed by an intermediate certificate missing from the server configuration. While the third reason is addressed at the server-side, the first two reasons are addressed by implementing a custom \textit{TrustManager}, an Android API, in the app. Implementation mistakes in the custom \textit{TrustManager} lead to vulnerabilities (e.g., trust all CAs) that can lead to Man-In-The-Middle attacks. 
    
    \item \textit{Unpinned Certificates.} Apps installed in a device trust all CAs that are pre-configured with the device. App developers can further restrict the CAs that the app will trust by pinning a set of trusted CAs to the app. The app then trusts only the pinned CAs and not any other CA, including the ones trusted by the device. Although not mandatory, certificate pinning is good security practice. However, they should be used with care as they can hamper usability due to communication failure because of outdated certificates as a result of changes to the server configuration.
\end{itemize}

Twelve of the 24 apps had at least one unprotected component vulnerability. We found several manifestations of this vulnerability. For example, apps used a third-party library that had an unprotected broadcast receiver that could write to shared preferences. Malicious apps could potentially exploit this vulnerability to execute data injection attacks. In other instances, apps defined an unprotected activity that could access Bluetooth and location. As a result, malicious apps could potentially exploit this activity to get access to privileged features without having the necessary permissions. Furthermore, few apps defined an activity that could share exposure-related diagnostic information with a remote server. Since this activity was exported without restriction, malicious apps could potentially exploit this vulnerability to communicate with the remote server. \textit{Apps must protect these components by making them private to the app. If these apps need to share these components with the underlying system, then they must protect the components with system permissions.}

All apps, except two, chose to use an insecure PRNG to generate the random identifiers used to identify devices where the app is installed anonymously. Choosing an insecure PRNG over a more secure PRNG allows malicious actors to potentially predict the random identifier more easily, which could be used to create duplicate identifiers and hence create erroneous or fake exposure-related data entries. \textit{Therefore, all the selected apps must use the \textit{SecureRandom} package to generate random identifiers}. 

Seven of the 24 apps use MD5 or SHA-1, which are weak hashing algorithms. Choosing a weak hashing algorithm for contact tracing apps is problematic since these apps are expected to provide strong privacy and security guarantees. \textit{A strong hashing algorithm is the most basic requirement with which the app can ensure the integrity of the information it stores}. 

A sixth of the 24 apps, that is, four apps allow users to back up the app’s data without rooting the device. This is insecure for GAEN apps since they store a user's and their contacts’ exposure-related data locally in the app. A malicious user with access to the device but without root access can get access to this data and misuse it. \textit{Apps that have this feature should disable it by setting the \textit{allowbackup} attribute in their manifest file to \textit{false}}.

Five of the 24 apps chose to use certificate pinning instead of relying on the device's list of trusted CAs. While certificate pinning provides additional protection against MITM attacks, most apps choose not to use it, possibly because of potential connection failures due to the pinned certificates becoming outdated. In such situations, the only way to restore the app is by pushing a software update, which the users must install. Temporary connection failures are not ideal and could render an app useless. \textit{However, in the context of the GAEN framework, where communication with remote servers is minimal\footnote{Users only connect with the remote server to upload a positive test result. Also, apps periodically connect with the remote server to check for positive cases.} and security is paramount, it is advisable for the apps to use certificate pinning to reduce the risk of an MITM attack}. Further, apps could pin backup certificates to prevent relying on only one pinned certificate. If one of them is outdated, the app can use the backups to connect to the server.    

MobSF reported an insecure SSL implementation vulnerability in exactly one app (the Arizona app) because the app was using a pinned self-signed certificate. Apps pinning self-signed certificates have benefits and limitations in terms of preventing MITM attacks. Consider a scenario where an app has pinned a certificate signed by a CA that has been compromised. In this situation, the app will need to be updated via a software update with a newly issued certificate. On the other hand, if the app had pinned a self-signed certificate, then the app only trusts that certificate and will not be affected if any other CA's certificate is compromised. However, using self-signed certificates are secure only if they are continuously monitored. In its absence, app developers may not know about a compromised certificate, and the app's communication with the compromised server will continue unknowingly. However, actively maintaining self-signed certificates is more cumbersome than using certificates from a trusted CA. Trusted CAs can revoke compromised certificates used by a server to stop communication between the app with the pinned certificate and the server, protecting the user from further harm. \textit{Therefore, in the general case, it is more secure to pin certificates signed by trusted CAs instead of self-signing them}. 

\section{Discussion}

\subsection{Observations on the apps}
The results show that contact tracing apps in Android based on the GAEN framework are over-privileged. Further, in Android, the apps are granted permission to use privileged system features at install time. As a result, users have less control over granting permission to these apps at runtime. This is concerning since these apps could potentially be used for tasks other than contact tracing, such as mass surveillance. Therefore, the apps must collect minimal information and use the least privileges.

The GAEN framework was developed to help create apps that preserve user privacy. However, our analysis shows that all apps violate their own privacy policy due to several potential reasons, such as developer oversight and developers’ lacking domain knowledge and awareness of the underlying platform.

Oversight is concerning but understandable since contact tracing apps were developed hurriedly to tackle the challenges of a growing pandemic. However, oversight in this context leads to a lack of transparency and credibility. It exacerbates the skepticism that the general public has towards contact tracing apps and hampers their widespread adoption. Weak adoption is not desirable since contact tracing apps if used effectively, are a vital tool to contain the pandemic. \textit{If developer oversight is the reason for privacy violations, then there is a need to develop tools and techniques that help developers write privacy policies that are consistent with their app’s behavior and vice versa}. 

Few apps in our study missed mentioning in their privacy policy all the personally identifiable information that they collected in their apps. One possible reason for this is a lack of domain knowledge in developers. In this context, privacy research efforts should focus on developing methods to help identify domain-specific personally identifiable information. Contact tracing apps have features that could be misused to violate users’ privacy. \textit{Therefore, it is crucial to accurately identify the information that the apps collect so users can make informed decisions about the privacy implications of using the apps}. 

The GAEN framework made it easier for healthcare providers to create apps for effective contact tracing without the need to know the details of the underlying platform \cite{gao}. This is also evident from certain privacy violations, which could have been avoided if developers had known about the underlying platform behavior. \textit{Therefore, existing research in app security and privacy should focus on developing methods and tools to assist less experienced developers gain the necessary knowledge to avoid privacy violations}.

The selected apps had vulnerabilities that are well known in the app development community. In fact, all four vulnerabilities that were discovered are part of the OWASP top 10 \cite{owasp10}, a popular set of guidelines for developing secure mobile and web apps. The apps had these vulnerabilities despite the focus on ensuring that the apps are secure and preserve user privacy. This suggests that the app developers did not have the experience to avoid these mistakes, or they did not have access to tools to help them prevent these vulnerabilities effectively. The latter is less likely since app development IDEs such as Android Studio have support for detecting and preventing such vulnerabilities during development. Therefore, it is more likely that the states did not allocate the resources to recruit developers with sufficient experience in mobile app development. \textit{While it is understandable that states need to prioritize their resources to tackle a pandemic, they should have planned better before developing and deploying apps with long-term consequences for user privacy and security}.  

\subsection{Observations on MobSF}

\begin{table*}[ht]
    \centering
    \resizebox{\textwidth}{!}{\begin{tabular}{l|l|l|l|l|l|l}
          & \multicolumn{5}{c}{False Positive Vulnerability} & \\
         \hline
         App & SQL Injection & Unprotected Component & Cleartext Storage & Log Sensitive & Janus Signature & \# False Positives\\
          & & & & Data & Vulnerability & per app\\
         \hline
         Alabama & Y & Y & Y & Y & Y & 5\\
         Arizona & Y & Y & N & Y & Y & 4\\
         California & Y & Y & Y & Y & Y & 5\\
         Colorado & Y & Y & Y & Y & Y & 5\\
         Connecticut & Y & Y & Y & Y & Y & 5\\
         Delaware & Y & Y & Y & Y & Y & 5\\
         DC & Y & Y & Y & Y & Y & 5\\
         Guam & Y & Y & Y & Y & Y & 5\\
         Hawaii & Y & Y & Y & Y & Y & 5\\
         Louisiana & Y & Y & Y & Y & Y & 5\\
         Maryland & Y & Y & Y & Y & Y & 5\\
         Michigan & Y & Y & Y & Y & Y & 5\\
         Minnesota & Y & Y & Y & Y & Y & 5\\
         Nevada & Y & Y & Y & Y & Y & 5\\
         New Jersey & Y & Y & Y & Y & Y & 5\\
         New Mexico & Y & Y & Y & Y & Y & 5\\
         New York & Y & Y & Y & Y & Y & 5\\
         North Carolina & Y & Y & Y & Y & Y & 5\\
         North Dakota/Wyoming & Y & Y & Y & Y & Y & 5\\
         Penn & Y & Y & Y & Y & Y & 5\\
         Utah & Y & Y & Y & Y & Y & 5\\
         Virginia & N & Y & N & Y & Y & 3\\
         Washington & Y & Y & Y & Y & Y & 5\\
         Wisconsin & Y & Y & Y & Y & Y & 5\\
         \hline
         \# Apps per & 23 & 24 & 22 & 24 & 24 \\
         false positive & & & & & \\
    \end{tabular}}
    \vspace{2mm}
    \caption{False Positives reported by MobSF in every US-based GAEN app. The cells with Y indicate that the vulnerability in the corresponding column was falsely reported as a potential vulnerability by MobSF and N denotes MobSF did not report the vulnerability in the corresponding column.}
    \label{tab:false_pos}
\end{table*}

While investigating the potential vulnerabilities MobSF reported, we discovered that a few of them were false positives, that is, falsely reported as vulnerabilities. We report and discuss them to help tool developers like MobSF improve their tools. 

Briefly, MobSF reported a total of five false positives across the 24 apps. Further, MobSF reported five false positives for all apps except the Virginia app and the Arizona app, which had three and four false positives, respectively. We explain the false positives, their likely reasons, and suggestions on how to avoid them as follows:

\begin{itemize}
    \item The \textit{SQLInjection} vulnerability was falsely reported in 23 of the 24 apps. MobSF reported the vulnerability in apps using the \textit{execSQL} API to execute SQL queries. However, using \textit{execSQL} leads to SQLInjection only if the query string has user-supplied input with potentially malicious SQL. None of the queries reported as being potentially vulnerable to SQLInjection relied on user input. Hence, they were not vulnerable to SQLInjection. 
    
    \textit{Suggestion: MobSF should flag queries in \textit{execSQL} only if they rely on user-input and are not parameterized}.
    
    \item MobSF falsely reported an \textit{unprotected component vulnerability} in all 24 apps because it found components in these apps that were exported. However, these components were also protected by system permissions, that is, only the system could be granted access to these components. Furthermore, these permissions were defined as part of the GAEN framework and were made available only for the purpose of contact tracing to government healthcare authorities. Therefore, apps without necessary authorization cannot request these permissions and get access to the exported components. 
    
    \textit{Suggestion: MobSF should consider the system permissions defined in the GAEN framework during analysis to improve the detection of unprotected components}.
    
    \item MobSF reported that 22 of the 24 apps saved sensitive data such as usernames, passwords, and secret keys in clear text. Furthermore, MobSF claimed that 22 of the 24 apps logged sensitive data. On further inspection, we found the claims to be false. MobSF reported the apps because they were saving or logging string constants in files and the strings contained words such as "key" and "password," but these constants were not sensitive data. 
    
    \textit{Suggestion: Considering that none of the string constants with words like key and password were sensitive data in our sample of apps, MobSF should consider using other heuristics to identify sensitive data}.
    
    \item MobSF falsely reported a Janus signature vulnerability in all 24 apps. Janus is a system vulnerability that allows attackers to inject a DEX file into an APK file signed with the v1 signature scheme without affecting the signatures. The vulnerability can be exploited because an Android package can be a valid DEX file and APK file at the same time. However, this vulnerability is exploitable only if the app runs on an Android version lower than 7.0. Android developers fixed this vulnerability in version 7.0 and above. As a result, Android APKs or packages have to be signed with the v2 and v3 signature schemes. All 24 contact tracing apps we considered used the v2 and v3 signature schemes to sign their APKs. MobSF flagged the apps as potentially vulnerable to Janus because the apps were also signed with the v1 signature scheme to enable backward compatibility. We categorize this as a false positive since the app developers have no choice but to sign their apps with the v1 signature scheme to support Android versions less than 7.0. Moreover, the app developers are doing due diligence by signing the apps with v2 and v3 signature schemes along with the v1 scheme (for backward compatibility), which is the best they can do under the circumstance.   
    
    \textit{Suggestion: MobSF should flag apps as potentially vulnerable to Janus if they are signed only with the v1 signature scheme. For apps signed with v1,v2, and v3 signatures, MobSF should consider reporting a different label like an information label to inform the developers that while this cannot be fixed at the app stage, one should be aware that the v1 signature is vulnerable on Android versions less than 7. Therefore, apps might want to consider supporting only Android version 7.0 or more}.
\end{itemize}

In conclusion, MobSF reported more false positives than potential true positives w.r.t potential vulnerabilities in each app (see Tables \ref{tab:known_vulns} vs. \ref{tab:false_pos}), which shows that MobSF has a high false positive rate. This observation is consistent with prior research efforts to evaluate the effectiveness of security analysis tools for Android apps \cite{ranganath2020}. In general, static analysis tools like MobSF have a high false positive rate, which hampers their adoption and reduces their effectiveness. Due to a high false positive rate, MobSF users have to manually verify the warnings and issues reported, which reduces trust in the tool verdicts. Consequently, this hampers tool adoption and the overall effectiveness of the tool.   

\section{Related Work}

Since the onset of the pandemic, numerous proposals have been made to automate the contact tracing process \cite{ahmed2020, bell2020}. In this context, many contact tracing apps have been implemented and deployed worldwide. This has led to a plethora of efforts to survey the characteristics and evaluate the effectiveness of contact tracing apps \cite{samhi2021, cho2020, lanzing2021}. Several agencies worldwide have called for an independent assessment of the security and privacy risks posed by contact tracing apps for greater transparency and accountability \cite{gao}. In this context, researchers have evaluated the security and privacy of the contact tracing apps to understand if they pose any privacy risks to the users \cite{baumgartner2020, gvili2020}. In this section, we discuss the efforts that are most closely related to our work and how they are different.

Wen et al. \cite{wen2020} performed a systematic study of 41 contact tracing apps deployed on Android and iOS. They used program analysis to determine the APIs relevant for contact tracing and identify the information collected by the apps. Additionally, they performed a cross-platform comparison of apps available on Android and iOS. Their results show that some apps expose identifiable information that can enable fingerprinting of apps and tracking of specific users. Moreover, they observed that some apps exhibited inconsistencies across platforms, which led to different privacy implications across the platforms. Their effort was one of the first attempts to understand the privacy and security implications of contact tracing apps. Although their effort is related to our evaluation, we focus on different aspects. For example, in addition to vulnerability analysis, we also analyze the privacy policy of the apps, which their effort does not consider. 

Samhi et al. \cite{samhi2021} conducted an empirical evaluation of Android apps in Google Play related to COVID-19. The aim of their study was to broadly characterize the apps in terms of their purpose, intended users, complexity, development process, and potential security risks. While their focus was not on security and privacy, they observed that none of the apps they considered leaked sensitive data based on static analysis of the apps using tools FlowDroid and IccTA. However, recent efforts to measure the effectiveness of security analysis tools in Android have raised questions on the effectiveness of static analysis tools like FlowDroid and IccTA in detecting sensitive data leaks \cite{ranganath2020, kouliaridis2021}.        

Hatmian et al. \cite{hatamian2021} analyzed the privacy and security performance of 28 contact tracing apps available on the Android platform in May-June 2020. They analyzed the permissions used by the apps and the potential vulnerabilities in them. Further, they measured the coverage of the privacy policy of the apps w.r.t the privacy principles outlined in the General Data Protection Regulation (GDPR) laws \cite{gdpr}. Our work is different from theirs in a number of ways. First, we focus on official apps developed by the US states based on the GAEN framework. None of the four US apps in the study conducted by Hatmian et al. are based on the GAEN framework. Second, instead of measuring the coverage of the apps' privacy policies w.r.t the privacy principles, we analyze if the apps' privacy policies are consistent with their source code, that is, apps are not violating their own privacy policies. Third, we critically analyze the verdicts reported by tools such as MobSF instead of reporting them as is. For example, Hatmian et al. report, based on MobSF's analysis, that all apps they considered log sensitive information. In our evaluation, MobSF also flagged all apps as logging sensitive information. However, after manually verifying the verdict, we discovered this was a false positive for all apps. 

In November 2020, Baumgartner et al. \cite{baumgartner2020} demonstrated that the GAEN framework's design is vulnerable to profiling and de-anonymizing infected persons and relay-based wormhole attacks that are capable of generating fake contacts to derail the contact tracing process. They claimed that if the vulnerabilities are not addressed, then all apps based on the framework will be vulnerable. Instead of analyzing the GAEN framework's design, in this paper, we are analyzing the apps based on the GAEN framework from the perspective of whether the apps comply with their own privacy policy and if they contain vulnerabilities that manifest due to known implementation bugs and incorrect configurations.   

Ang et al. \cite{ang2021} reviewed the security and privacy of 70 contact tracing apps one year after the pandemic. They statically analyzed the apps using MobSF for vulnerabilities based on threat scenarios they identified for contact tracing apps. Additionally, they reported data trackers embedded in apps that can potentially violate privacy since data collected by the trackers can be used without the users' consent. However, their privacy analysis does not include any analysis of an app's privacy policy. The set of apps in the evaluation by Ang et al. includes 20 apps that we considered in our assessment. However, the results of our analysis differ significantly. For example, 80\% of the apps they considered stored sensitive information in cleartext. On the other hand, we found this to be a false positive for 20 of the 24 apps we considered. Similarly, we discovered vulnerabilities in our analysis (e.g., data backup) that are not reported in Ang et al. Further, Ang et al. used dynamic analyzers such as VirusTotal \cite{vtotal} to detect the presence of malware in the contact tracing apps. However, malware detection tools such as VirusTotal do not accurately detect the presence of malware as they often falsely identify Potentially Unwanted Programs (PUPs) as malware \cite{ranganath2020}.   

Kouliaridis et al. \cite{kouliaridis2021} investigated all official contact tracing apps deployed by European countries as of Feb 2, 2021. They analyzed the apps both statically and dynamically. Static analysis included sensitive permissions and API calls, third-party trackers, and known vulnerabilities and configurations that affect app security based on the Common Weakness Enumerations (CWEs) \cite{cwe} and Common Vulnerabilities and Exposures (CVEs) \cite{cve}. Dynamic analysis involved instrumenting the app's source code, verifying if the app uses location and Bluetooth services at runtime, and monitoring network traffic. The evaluation in our paper differs from Kouliaridis et al. in three distinct ways. First, we considered a different set of apps. Second, we analyzed the apps' privacy policies to determine if they are consistent with their encoded behavior. Third, we did not dynamically analyze the apps. Further, there are notable differences in our static analysis observations. For example, Kouliaridis et al. report that two-thirds of their apps, which included GAEN apps, had a potential SQL injection vulnerability based on MobSF's analysis. However, we observed that MobSF falsely reported SQL injection as a vulnerability in 23 of the 24 GAEN apps we considered.

\section{Caveats}

In vulnerability analysis, we considered vulnerabilities in Ghera and reported by MobSF. Since we did not cover vulnerabilities outside these sources, it is possible that they existed in the apps but went unreported. Consequently, our vulnerability analysis may not be comprehensive. App developers reading this report must take steps to fix the reported vulnerabilities and perform further analysis to ensure that other vulnerabilities not reported here do not exist in their apps. 

We confirmed the potential vulnerabilities reported by static analysis tools by manually examining them. However, we did not build malicious applications to exploit the vulnerabilities. Therefore, we do not know to what extent the vulnerabilities are exploitable.

The results reported in this study are limited to the set of apps considered or the GAEN apps in general. Therefore, they should not be generalized for other contact tracing apps, especially apps not based on the GAEN framework.

\section{Conclusion}

In this paper, we conducted a systematic investigation of 24 contact tracing apps based on the GAEN framework in the US. All the apps were implemented and deployed by the official health departments of the respective US states. We discovered that the considered apps are over-privileged, they violate their own privacy policies, and contain vulnerabilities that can be exploited by malicious users to cause harm to the app's users.

\subsection*{Call For Action}
While there have been previous efforts to evaluate the contact tracing apps for privacy violations and vulnerabilities, none of them have focused on the consistency of the apps' privacy policies w.r.t to their encoded behavior. Although there are similarities between our effort and existing efforts in terms of vulnerability analysis of contact tracing apps, the results differ markedly. Our results show that few vulnerabilities reported as potential vulnerabilities in related evaluations are false positives. For example, several existing research efforts have reported the Janus vulnerability, which we reported as a false positive, as a true positive in their results. \textit{Therefore, this raises the question if efforts to study the privacy and security of contact tracing apps are reporting vulnerabilities that may not manifest in reality}. Reporting false positives as potential true positives may erode the public's trust in contact tracing apps and eventually lead to reduced adoption, which may ultimately weaken efforts to contain the pandemic. Therefore, \textit{there is a need for researchers to continuously evaluate the security and privacy of contact tracing apps to reproduce and verify the results}. 

\section{Ackowledgements}

We wish to thank Minqi Shi, Taylor Giles, Soroush Semerkant,  Mihir Madhira, Jeffrey Jiminez, Colin Ruan, and Patrick Wszeborowski, undergraduate students in the Department of Computer Science at Stony Brook University for assisting with data collection.

\bibliographystyle{IEEEtran}
\bibliography{main}

\begin{thebibliography}{10}
\providecommand{\url}[1]{#1}
\csname url@samestyle\endcsname
\providecommand{\newblock}{\relax}
\providecommand{\bibinfo}[2]{#2}
\providecommand{\BIBentrySTDinterwordspacing}{\spaceskip=0pt\relax}
\providecommand{\BIBentryALTinterwordstretchfactor}{4}
\providecommand{\BIBentryALTinterwordspacing}{\spaceskip=\fontdimen2\font plus
\BIBentryALTinterwordstretchfactor\fontdimen3\font minus
  \fontdimen4\font\relax}
\providecommand{\BIBforeignlanguage}[2]{{%
\expandafter\ifx\csname l@#1\endcsname\relax
\typeout{** WARNING: IEEEtran.bst: No hyphenation pattern has been}%
\typeout{** loaded for the language `#1'. Using the pattern for}%
\typeout{** the default language instead.}%
\else
\language=\csname l@#1\endcsname
\fi
#2}}
\providecommand{\BIBdecl}{\relax}
\BIBdecl

\bibitem{klinkenberg2006}
D.~Klinkenberg, C.~Fraser, and H.~Heesterbeek, ``The effectiveness of contact
  tracing in emerging epidemics,'' \emph{PloS one}, vol.~1, no.~1, p. e12,
  2006.

\bibitem{jiang2022}
T.~Jiang, Y.~Zhang, M.~Zhang, T.~Yu, Y.~Chen, C.~Lu, J.~Zhang, Z.~Li, J.~Gao,
  and S.~Zhou, ``A survey on contact tracing: the latest advancements and
  challenges,'' \emph{ACM Transactions on Spatial Algorithms and Systems
  (TSAS)}, vol.~8, no.~2, pp. 1--35, 2022.

\bibitem{anglemyer2020}
A.~Anglemyer, T.~H. Moore, L.~Parker, T.~Chambers, A.~Grady, K.~Chiu, M.~Parry,
  M.~Wilczynska, E.~Flemyng, and L.~Bero, ``Digital contact tracing
  technologies in epidemics: a rapid review,'' \emph{Cochrane Database of
  Systematic Reviews}, no.~8, 2020.

\bibitem{ahmed2020}
N.~Ahmed, R.~A. Michelin, W.~Xue, S.~Ruj, R.~Malaney, S.~S. Kanhere,
  A.~Seneviratne, W.~Hu, H.~Janicke, and S.~K. Jha, ``A survey of covid-19
  contact tracing apps,'' \emph{IEEE access}, vol.~8, pp. 134\,577--134\,601,
  2020.

\bibitem{akinbi2021}
A.~Akinbi, M.~Forshaw, and V.~Blinkhorn, ``Contact tracing apps for the
  covid-19 pandemic: a systematic literature review of challenges and future
  directions for neo-liberal societies,'' \emph{Health Information Science and
  Systems}, vol.~9, no.~1, pp. 1--15, 2021.

\bibitem{rowe2020}
F.~Rowe, ``Contact tracing apps and values dilemmas: A privacy paradox in a
  neo-liberal world,'' \emph{International Journal of Information Management},
  vol.~55, p. 102178, 2020.

\bibitem{hassandoust2021}
F.~Hassandoust, S.~Akhlaghpour, and A.~C. Johnston, ``Individuals’ privacy
  concerns and adoption of contact tracing mobile applications in a pandemic: A
  situational privacy calculus perspective,'' \emph{Journal of the American
  Medical Informatics Association}, vol.~28, no.~3, pp. 463--471, 2021.

\bibitem{martin2020}
T.~Martin, G.~Karopoulos, J.~L. Hern{\'a}ndez-Ramos, G.~Kambourakis, and
  I.~Nai~Fovino, ``Demystifying covid-19 digital contact tracing: A survey on
  frameworks and mobile apps,'' \emph{Wireless Communications and Mobile
  Computing}, vol. 2020, 2020.

\bibitem{gao}
U.~S. G.~A. Office, ``Benefits and challenges of smartphone applications to
  augment contact tracing,'' https://www.gao.gov/products/gao-21-104622, Sept
  2021.

\bibitem{tracetogether2020}
T.~TraceTogether, ``How does tracetogether work,'' 2020.

\bibitem{gupta2020}
R.~Gupta, M.~Bedi, P.~Goyal, S.~Wadhera, and V.~Verma, ``Analysis of covid-19
  tracking tool in india: case study of aarogya setu mobile application,''
  \emph{Digital Government: Research and Practice}, vol.~1, no.~4, pp. 1--8,
  2020.

\bibitem{vaudenay2020}
S.~Vaudenay, ``Centralized or decentralized? the contact tracing dilemma,''
  \emph{Cryptology ePrint Archive}, 2020.

\bibitem{li2020}
T.~Li, C.~Faklaris, J.~King, Y.~Agarwal, L.~Dabbish, J.~I. Hong \emph{et~al.},
  ``Decentralized is not risk-free: Understanding public perceptions of
  privacy-utility trade-offs in covid-19 contact-tracing apps,'' \emph{arXiv
  preprint arXiv:2005.11957}, 2020.

\bibitem{googleGuide}
Google, ``Exposure notifications implementation guide,''
  \url{https://developers.google.com/android/exposure-notifications/implementation-guide},
  Feb. 2022.

\bibitem{appleGuide}
Apple, ``Enexposureconfiguration,''
  \url{https://developer.apple.com/documentation/exposurenotification/enexposureconfiguration},
  Feb. 2022.

\bibitem{altmann2020}
S.~Altmann, L.~Milsom, H.~Zillessen, R.~Blasone, F.~Gerdon, R.~Bach,
  F.~Kreuter, D.~Nosenzo, S.~Toussaert, J.~Abeler \emph{et~al.},
  ``Acceptability of app-based contact tracing for covid-19: Cross-country
  survey study,'' \emph{JMIR mHealth and uHealth}, vol.~8, no.~8, p. e19857,
  2020.

\bibitem{googleApps}
Google, ``Publicly-available exposure notifications apps,''
  \url{https://developers.google.com/android/exposure-notifications/apps}, Feb.
  2022.

\bibitem{Sufatrio2015}
Sufatrio, D.~J.~J. Tan, T.-W. Chua, and V.~L.~L. Thing, ``Securing android: A
  survey, taxonomy, and challenges,'' \emph{ACM Comput. Surv.}, pp.
  58:1--58:45, 2015.

\bibitem{li2017}
L.~Li, T.~F. Bissyand{\'e}, M.~Papadakis, S.~Rasthofer, A.~Bartel, D.~Octeau,
  J.~Klein, and L.~Traon, ``Static analysis of android apps: A systematic
  literature review,'' \emph{Information and Software Technology}, vol.~88, pp.
  67--95, 2017.

\bibitem{ranganath2020}
V.-P. Ranganath and J.~Mitra, ``Are free android app security analysis tools
  effective in detecting known vulnerabilities?'' \emph{Empirical Software
  Engineering}, vol.~25, no.~1, pp. 178--219, 2020.

\bibitem{mobsf}
A.~Abraham, Magaofei, M.~Dobrushin, and V.~Nadal, ``Mobsf github,''
  \url{https://github.com/MobSF/Mobile-Security-Framework-MobSF}, Feb. 2022.

\bibitem{androguard}
A.~Desnos, G.~Gueguen, and S.~Bachmann, ``Androguard,''
  \url{https://androguard.readthedocs.io/en/latest/}, Feb. 2022.

\bibitem{lu2012}
L.~Lu, Z.~Li, Z.~Wu, W.~Lee, and G.~Jiang, ``Chex: statically vetting android
  apps for component hijacking vulnerabilities,'' in \emph{Proceedings of the
  2012 ACM conference on Computer and communications security}, 2012, pp.
  229--240.

\bibitem{watanabe2017}
T.~Watanabe, M.~Akiyama, F.~Kanei, E.~Shioji, Y.~Takata, B.~Sun, Y.~Ishi,
  T.~Shibahara, T.~Yagi, and T.~Mori, ``Understanding the origins of mobile app
  vulnerabilities: A large-scale measurement study of free and paid apps,'' in
  \emph{2017 IEEE/ACM 14th International Conference on Mining Software
  Repositories (MSR)}.\hskip 1em plus 0.5em minus 0.4em\relax IEEE, 2017, pp.
  14--24.

\bibitem{mitra2017}
J.~Mitra and V.-P. Ranganath, ``Ghera: A repository of android app
  vulnerability benchmarks,'' in \emph{Proceedings of the 13th International
  Conference on Predictive Models and Data Analytics in Software Engineering},
  2017, pp. 43--52.

\bibitem{ghafari2017}
M.~Ghafari, P.~Gadient, and O.~Nierstrasz, ``Security smells in android,'' in
  \emph{2017 IEEE 17th international working conference on source code analysis
  and manipulation (SCAM)}.\hskip 1em plus 0.5em minus 0.4em\relax IEEE, 2017,
  pp. 121--130.

\bibitem{pew2015permissions}
P.~R. Center, ``An analysis of android app permissions,''
  \url{https://www.pewresearch.org/internet/2015/11/10/an-analysis-of-android-app-permissions/},
  Nov 2015.

\bibitem{sarma2012}
B.~P. Sarma, N.~Li, C.~Gates, R.~Potharaju, C.~Nita-Rotaru, and I.~Molloy,
  ``Android permissions: a perspective combining risks and benefits,'' in
  \emph{Proceedings of the 17th ACM symposium on Access Control Models and
  Technologies}, 2012, pp. 13--22.

\bibitem{androidPerms}
Google, ``Android app permissions,''
  \url{https://developer.android.com/guide/topics/permissions/overview}, May
  2022.

\bibitem{prakash2022}
A.~V. Prakash and S.~Das, ``Explaining citizens’ resistance to use digital
  contact tracing apps: A mixed-methods study,'' \emph{International Journal of
  Information Management}, vol.~63, p. 102468, 2022.

\bibitem{fingerprintPerms}
Google, ``Manifest permissions in android,''
  \url{https://developer.android.com/reference/android/Manifest.permission#USE_FINGERPRINT},
  Jul. 2022.

\bibitem{owasp10}
Owasp, ``Owasp top 10,'' \url{https://owasp.org/www-project-mobile-top-10/},
  Jul. 2022.

\bibitem{bell2020}
J.~Bell, D.~Butler, C.~Hicks, and J.~Crowcroft, ``Tracesecure: Towards privacy
  preserving contact tracing,'' \emph{arXiv preprint arXiv:2004.04059}, 2020.

\bibitem{samhi2021}
J.~Samhi, K.~Allix, T.~F. Bissyand{\'e}, and J.~Klein, ``A first look at
  android applications in google play related to covid-19,'' \emph{Empirical
  Software Engineering}, vol.~26, no.~4, pp. 1--49, 2021.

\bibitem{cho2020}
H.~Cho, D.~Ippolito, and Y.~W. Yu, ``Contact tracing mobile apps for covid-19:
  Privacy considerations and related trade-offs,'' \emph{arXiv preprint
  arXiv:2003.11511}, 2020.

\bibitem{lanzing2021}
M.~Lanzing, ``Contact tracing apps: an ethical roadmap,'' \emph{Ethics and
  information technology}, vol.~23, no.~1, pp. 87--90, 2021.

\bibitem{baumgartner2020}
L.~Baumg{\"a}rtner, A.~Dmitrienko, B.~Freisleben, A.~Gruler, J.~H{\"o}chst,
  J.~K{\"u}hlberg, M.~Mezini, R.~Mitev, M.~Miettinen, A.~Muhamedagic
  \emph{et~al.}, ``Mind the gap: Security \& privacy risks of contact tracing
  apps,'' in \emph{2020 IEEE 19th international conference on trust, security
  and privacy in computing and communications (TrustCom)}.\hskip 1em plus 0.5em
  minus 0.4em\relax IEEE, 2020, pp. 458--467.

\bibitem{gvili2020}
Y.~Gvili, ``Security analysis of the covid-19 contact tracing specifications by
  apple inc. and google inc.'' \emph{Cryptology ePrint Archive}, 2020.

\bibitem{wen2020}
H.~Wen, Q.~Zhao, Z.~Lin, D.~Xuan, and N.~Shroff, ``A study of the privacy of
  covid-19 contact tracing apps,'' in \emph{International Conference on
  Security and Privacy in Communication Systems}.\hskip 1em plus 0.5em minus
  0.4em\relax Springer, 2020, pp. 297--317.

\bibitem{hatamian2021}
M.~Hatamian, S.~Wairimu, N.~Momen, and L.~Fritsch, ``A privacy and security
  analysis of early-deployed covid-19 contact tracing android apps,''
  \emph{Empirical software engineering}, vol.~26, no.~3, pp. 1--51, 2021.

\bibitem{gdpr}
E.~Union, ``General data protection regulation,''
  https://gdpr.eu/what-is-gdpr/, 2022.

\bibitem{ang2021}
V.~Ang and L.~K. Shar, ``Covid-19 one year on--security and privacy review of
  contact tracing mobile apps,'' \emph{IEEE Pervasive Computing}, vol.~20,
  no.~4, pp. 61--70, 2021.

\bibitem{vtotal}
H.~Systemas, ``Virustotal,'' https://www.virustotal.com/gui/home/upload, 2022.

\bibitem{kouliaridis2021}
V.~Kouliaridis, G.~Kambourakis, E.~Chatzoglou, D.~Geneiatakis, and H.~Wang,
  ``Dissecting contact tracing apps in the android platform,'' \emph{Plos one},
  vol.~16, no.~5, p. e0251867, 2021.

\bibitem{cwe}
Mitre, ``Common weakness enumeration,'' https://cwe.mitre.org/, 2022.

\bibitem{cve}
``Common vulnerabilities and exposures,'' https://cve.mitre.org/, 2022.

\end{thebibliography}

\end{document}